\documentclass[twocolumn,noshowpacs,nofootinbib,10pt]{revtex4-1}

\usepackage{float,xcolor,upgreek,yfonts,tikz}
\usepackage{color}
\usepackage{graphicx}
\usepackage{graphicx,float}\usepackage[all]{xy}
\usepackage{amsmath,upgreek}
 
\usepackage{amssymb}

\usepackage{alphabeta}
\usepackage{textgreek}

\newcommand{\beq}{\begin{eqnarray}}
\newcommand{\eeq}{\end{eqnarray}}
\newcommand{\be}{\begin{equation}}
\newcommand{\ee}{\end{equation}}

\newcommand{\bea}{\begin{eqnarray}}
\newcommand{\eea}{\end{eqnarray}}

\newcommand{\ba}{\begin{eqnarray}}
\newcommand{\ea}{\end{eqnarray}}
\newcommand{\clt}{\textcolor{black}}

\usepackage[colorlinks,hyperindex,unicode]{hyperref}
\definecolor{green1}{RGB}{0,128,0} 
\hypersetup{hidelinks,backref=true,pagebackref=true,hyperindex=true,colorlinks=true,breaklinks=true,urlcolor= blue}
\hypersetup{%
  colorlinks = true,
  linkcolor  = blue,
  citecolor = cyan,
}
\usepackage{bookmark,textgreek}
\usepackage{hyperref,color,xcolor}
\hypersetup{hidelinks,hyperindex=true,colorlinks=true,breaklinks=true,urlcolor= blue}
\hypersetup{%
  colorlinks = true,
  linkcolor  = blue
}
\newcommand\orcidroldao{{\href{https://orcid.org/0000-0003-3978-532X}{\orcidicon}}}
\newcommand{\orcidicon}{%
	\begin{tikzpicture}
	\draw[lime, fill=lime] (0,0)
		circle [radius=0.16]
		node[white] {{\fontfamily{qag}\selectfont \tiny ID}};
	\draw[white, fill=white] (-0.0625,0.095)
		circle [radius=0.007];
	\end{tikzpicture}	\hspace{-2mm}
}

\begin{document}
\title{AdS graviton stars and differential configurational entropy}

\author{Roldao da Rocha\orcidroldao\!\!}
\affiliation{Center of Mathematics,  Federal University of ABC, 09210-580, Santo Andr\'e, Brazil.}
\email{roldao.rocha@ufabc.edu.br}
\begin{abstract}
AdS graviton stars are studied in the differential configurational entropy setup, as solutions of the effective Einstein field equations that backreact to compactification. With the critical central density of AdS graviton stars, the differential configurational entropy is derived and computed, presenting global minima for a wide range of stellar mass magnitude orders. It indicates insular domains of configurational stability for AdS graviton stars near astrophysical neutron star densities. Other relevant features are also reported. 

 \end{abstract}
\maketitle

\section{Introduction}

The entropy of information has been employed to explore and quantify features of physical systems, since after Gibbs and Boltzmann formalized the idea of entropy in statistical mechanics. 
When regarding any physical system with randomness, evolving and undergoing chance fluctuations, 
the statistical point of view of physical processes was complemented with Shannon information entropy \cite{Shannon:1948zz}. Although several applications in several areas of physics have been developed since the 1950s, a vast spectrum of relevant developments and new applications of information entropy have been boosted again in the last decade. Analogously to the Boltzmann--Gibbs entropy, Shannon information entropy also quantifies the level of disorder inherent to physical systems. When continuous physical systems are taken into account, the most appropriate information entropy measure is the differential configurational entropy (DCE) \cite{Gleiser:2018kbq}, which can be engendered by the localized and integrable density that underlies any physical solution of Euler--Lagrange equations \cite{Gleiser:2011di,Gleiser:2012tu,Bernardini:2016hvx}. 
Critical values of the DCE correspond to physical states that are more prevalent and dominant, among all possible states that can be occupied by the system. A large variety of physical systems have been investigated with the tools of the DCE. Phenomenological aspects of gauge/gravity dualities, using the DCE, were addressed in Refs. \cite{Karapetyan:2018oye,Bernardini:2018uuy,Karapetyan:2018yhm,daRocha:2021imz,Ferreira:2019nkz,Karapetyan:2020yhs,Braga:2018fyc,Braga:2017fsb,Braga:2020hhs,Braga:2020myi,Ferreira:2020iry}.
Phase transitions driven by the critical values of the DCE have been also identified for several physical systems \cite{Bernardini:2019stn,Sowinski:2015cfa}. The DCE was employed to study AdS black holes \cite{Braga:2016wzx,Braga:2019jqg,Braga:2020opg,Lee:2017ero}, in a coherent setup that corroborates to the Hawking-Page phase transition, and also to study compact stars and their stability \cite{Casadio:2016aum,Gleiser:2013mga,Fernandes-Silva:2019fez,Gleiser:2015rwa,Gleiser:2018jpd,Lee:2017ero}.  
Other physical applications involving the DCE also explore topological defects  
\cite{Bazeia:2018uyg,Bazeia:2021stz,Correa:2016pgr,Cruz:2019kwh}. All these important developments take into account the concept of minimum message length, as a Bayesian information procedure that underlies the DCE concept. When physical models are equivalent in their measure of fit-accuracy to the observed data, the one generating the most concise and condensed information in data is more likely to be correct, in the lossless encoding regime.
Therefore critical values of the DCE correspond to physical states corresponding to the most compressed 
information in momentum space.

In this work, the DCE of self-gravitating AdS${}_4$ graviton stars will be computed and discussed. 
One calls a graviton star a self-gravitating system that is constituted by massless supergravity bosonic fields, where the temperature is sufficiently high to induce backreaction to the star geometric background. The solutions of the field equations that will be studied here have some qualitative similarities to radiation stars \cite{Arsiwalla:2010bt,Page:1985em}, which consist of equilibrium configurations of self-gravitating massless radiation in asymptotically AdS space.
Ref. \cite{Witten:1998zw} showed that both radiation and graviton stars can also portray superheated phases of
some large-$N$ Yang--Mills theories, that go through deconfinement, which emulates the Hawking--Page phase transition between a gas of gravitons and a black
hole. When the microcanonical ensemble prescription is taken into account, 
driving energy to heat the gas above the Hawking-Page temperature. The superheated gas phase 
 carries on up to gravitational
backreaction to set in \cite{Arsiwalla:2010bt}.
To compute the backreaction of the gas of gravitons, one must solve Einstein field equations coupled to the stress-energy-momentum tensor of a radiation fluid. Therefore graviton stars can be studied with the same apparatus that is used to approach radiation stars \cite{Page:1985em}, as long as the equation of state is reformulated to make the density encompass the different graviton polarizations, on effective AdS$_{4}$ backgrounds under spontaneous compactification \cite{Arsiwalla:2010bt}. In this way, the equations of motion in bosonic supergravity can be completed by a
radiation fluid term \cite{Martin:1999iy}.

This paper is organized as follows: Sec. \ref{17} is devoted to present AdS${}_4$ graviton stars. Metric solutions of the effective Einstein field equations are deformed by a dilaton field that regulates the symmetry of static configurations under the Freund--Rubin ansatz, as well as encompasses the backreaction to the geometric background. The 4-dimensional equilibrium configurations of AdS${}_4$ graviton stars are then represented by their density and an equation of state, also involving the Misner--Sharp--Hernandez mass function. 
In Sec. \ref{2s}, the density is the main ingredient, as an integrable scalar field that is spatially localized and describes important features of graviton stellar configurations, whose DCE is then computed and discussed. The critical central density values of AdS${}_4$ graviton stars have corresponding global minima of the DCE, that indicate insular domains of configurational stability. Subsequent features and analysis are reported, also in Sec. \ref{conclu}.

\section{AdS${}_4$ graviton stars}
\label{17}

Spontaneous compactification of bosonic supergravity consists of field equations solutions for which the ground state is characterized by a product space between a 4-dimensional spacetime and a compact 7-dimensional space \cite{Freund:1980xh,Duff:1986hr}. AdS${}_4$ graviton stars can be emulated in a background compactification setup. The action regards Einstein--Hilbert gravity coupled to a Yang--Mills field strength, 
\beq
S = {1\over 16\pi{}{G}} \int d^{\scalebox{.56}{$11$}}x \sqrt{{}{-g}} \left(
{}{R} -{1\over 2}F^2\right),
\eeq 
where hereon greek indexes run from $1,\ldots,11$; the Ricci scalar curvature is denoted by $R$, $g_{\mu\nu}$ denotes the metric, and $F^2=F_{\mu\nu\alpha\beta}F^{\mu\nu\alpha\beta}$, where the field strength $F_{\mu\nu\alpha\beta}= \partial_{[\mu}A_{\nu\alpha\beta]}$ is generated by the gauge field $A_{\nu\alpha\beta}$, providing a dynamical mechanism of spontaneous compactification \cite{Duff:1986hr,Witten:1981me}.
The equations of motion for the bosonic fields $g_{\mu\nu}$ and $A_{\nu\alpha\beta}$, \clt{respectively describing gravity and matter, } \!are given by 
\beq
\label{EOMSeleven1}
\!\!\!\!\!\!\!\!\!\!\!{}{R}_{\mu\nu} -{1\over 2}{}{g}_{\mu\nu}{}{R} &=& {1\over 6} F_{\mu\sigma\alpha\beta}F_{\nu}^{\sigma\alpha\beta} -{1\over 4}F^2{}{g}_{\mu\nu},\\
\!\!\!\!\!\!\!\!\nabla_{\alpha}\!\left(\sqrt{-g}F_{\alpha\nu\sigma\tau}\right)&=&-\frac{1}{1152}\epsilon_{\nu\sigma\tau\mu_1\ldots\mu_{8}}F^{\mu_1\ldots\mu_{4}}F^{\mu_{5}\ldots\mu_{8}},\label{EOMSeleven2}
\eeq
where $R_{\mu\nu}$ is the Ricci tensor and $\epsilon$ is a fully antisymmetric covariant constant tensor. \clt{This setup can be interpreted, in first-order formalism, as a pure gravity theory with torsion, when the $A_{\nu\alpha\beta}$ gauge invariance is regarded as a spacetime symmetry of pure gravity \cite{Bars:1983tc}}. The Freund--Rubin ansatz states that 
the field strength is proportional to the 4-dimensional volume form \cite{Freund:1980xh},  
\beq\label{vol}
F_{\mu\nu\alpha\beta} = \frac{3\lambda}{2} \epsilon_{\mu\nu\alpha\beta},\qquad m\in\mathbb{R}.\eeq
Hereon Latin indexes, regarding sans-serif typestyle quantities, refer to 4-dimensional quantities. 
 Eq. (\ref{vol}) makes \eqref{EOMSeleven2} to be trivially satisfied, whereas Eq. \eqref{EOMSeleven1} yields the product of the 7-sphere $S^7$ and an AdS${}_4$ spacetime, with Ricci tensor $R_{ab} =-\frac{3}{\lambda^2}g_{ab}$ and cosmological constant $\Lambda_4=-12\lambda^2$ \cite{Freund:1980xh,Duff:1986hr}. From the phenomenological point of view, $S^7$ has size near Planck scale, being negligible and undetectable at currently available energies at running experiments \cite{Salam:1981xd,Cremmer:1978km}.
\clt{The Freund--Rubin ansatz provides 
the $S^7$ internal geometry that translates into concrete aspects in the effective 4-dimensional theory. The Englert--Brout--Higgs--Guralnik--Hagen--Kibble mechanism describes the way how gauge vector bosons can acquire nonzero masses in the process of spontaneous symmetry breaking in electroweak theory \cite{Guralnik:1964eu,Englert:1964et,Higgs:1964ia}. This mechanism endows scalars and pseudoscalars to acquire non-zero vacuum expectation values, also allowing the spontaneous breakdown of gauge symmetries, supersymmetries, and discrete symmetries, like parity, charge conjugation, and their composition as well. The Freund--Rubin compactification procedure is also the only non-trivial compactification of any Kaluza--Klein type model that circumvents and avoids the inconsistencies of the Kaluza--Klein ansatz. }

 The 
AdS$_{4}\times$$S^7$ solution has metric
\beq\label{def}
ds^2 = \ell^2 \mathsf{g}_{ab}dx^adx^b + \lambda^2 d\Omega_7^2
\eeq where $\mathsf{g}_{ab}dx^adx^b$ is the ${{\rm AdS}_{4}}$ line element and the AdS${}_4$ radius $\ell$ satisfies $
\ell^2 = - {3\over \Lambda_4}$, whereas $d\Omega_7^2$ denotes the solid angle element of $S^7$. 
In the presence of a radiation fluid with 
 stress-energy-momentum tensor emulating that of a perfect fluid, 
\beq
T_{\mu\nu} = ({}{\rho} +{}{p}){}{u}_\mu
{}{u}_\nu +{}{p}{}{g}_{\mu\nu},
\eeq

 Eq. (\ref{EOMSeleven1}) can be promoted to the form 
\beq
\!\!\!\!\!\!\!\!\!\!\!\!{}{R}_{\mu\nu} \!-\!{1\over 2}{}{g}_{\mu\nu}{}{R} = {1\over 12} F_{\mu\sigma\alpha\beta}F_{\nu}^{\sigma\alpha\beta} \!-\!{1\over 4}F^2{}{g}_{\mu\nu}+
8\pi{}{G} T_{\mu\nu},
\eeq
with equation of state for radiation
\beq\label{eosr}
{p} = {\rho\over 10}.\eeq
The fluid velocity satisfies $g_{\mu\nu}{u}^\mu
{}{u}^\mu = -1$. Static configurations that preserve the $SO(8)$ symmetry of $S^7$ make its radius to vary according to a dilaton $\varphi$ \cite{Duff:1986hr,Witten:1981me}. It also encodes the backreaction to the graviton star geometrical background deforming the AdS$_{4}\times$$S^7$ metric (\ref{def}). Ref. \cite{Arsiwalla:2010bt} proposed the ansatz
\beq\label{ansa}
ds^2 = e^{-\varphi(x^c)} \mathsf{g}_{ab} (x^c)\clt{dx^adx^b} + \lambda^2 e^{2 \varphi(x^c)} d\Omega_7^2.
\eeq 
The 11-dimensional Newton constant, $\mathring{G}$, is related to the $4$-dimensional one by $
 {\mathring{G}}={\lambda^7 V_7G}$, where $V_7(\lambda^7)=\frac{16\pi^3}{105}\lambda^7$ is the volume of $S^7$. 
The 
 $4$-dimensional effective pressure and density take the form
\beq
\mathsf{p} = \frac{G}{\mathring{G}} e^{-7\varphi}{}{p},\qquad \mathsf{\uprho} = \frac{G}{\mathring{G}} 
e^{-7\varphi}{}{\rho}.\eeq

Identifying 
$\Lambda_4={F^2}/3$ yields Eqs. (\ref{EOMSeleven1}, \ref{EOMSeleven2}) to be equivalently expressed as $4$-dimensional 
equations \cite{Arsiwalla:2010bt}
\begin{widetext}
\beq\label{jr1}
\!\!\!\!\!\!\!\!\!\!\!\mathsf{R}_{ab} -{1\over 2}\mathsf{g}_{ab}\mathsf{R} +\Lambda_4 \mathsf{g}_{ab}&=&{63 \over 2}\left(\partial_a \varphi \partial_b \varphi - {1\over 2} 
\mathsf{g}_{ab} \partial_c\varphi \partial^c\varphi - V(\varphi)\mathsf{g}_{ab}\right) 
+ 8 \pi G\left[({\uprho}+\mathsf{p})\mathsf{u}_a \mathsf{u}_b + \mathsf{p} \mathsf{g}_{ab}\right],\\
 \Box\varphi &=& \frac{\partial V(\varphi)}{\partial r} - 8\pi G \mathsf{p},\label{jr2}
 \eeq
 \end{widetext}
where the potential that governs the dilaton reads 
\clt{\beq\label{pott}
V(\varphi) = -{\Lambda_4 \over 378}\left[\left(9e^{-21\varphi}- 21e^{-9\varphi} \right) - {4\over
21}\right].\eeq}
\!\!The effective AdS$_{4}$ graviton star solution can be now implemented when one considers a spherically symmetric ansatz for the $4$-dimensional metric \cite{Arsiwalla:2010bt}
\beq\label{ast1}
ds^2 =-e^{2\xi(r)-2\theta(r)} dt^2 + e^{2\theta(r)} dr^2 + r^2 d\Omega_{2}^2.
\eeq \clt{With the metric \eqref{ast1}, Eq. (\ref{jr2}) that governs the dilaton reads}
\beq
\!\!\!\!\!\!\!\!\!\!\!\!\frac{d^2 \varphi}{dr^2} \!+\! 
\left(\frac{d\xi}{dr}\!-\!2\frac{d\theta}{dr} \!+\! {2 \over r} \right) \frac{d \varphi}{dr} \!=\! e^{2\theta}\left(\frac{\partial V(\varphi)}{\partial r} \!-\! 8 \pi G \mathsf{p}\right),
\eeq
whereas \clt{the equation ruling the temporal ({\footnotesize{$tt$}}) component of Eq. (\ref{jr1})} yields 
\beq\label{fi1}
&&\!\!\!\!\!\!\! \left[(e^{2\theta}\!-\!1) \!+\! 2 r \frac{d\theta}{dr}\right]\!\frac{e^{-2\theta}}{r^2}
\!+\! {3\over \ell^2} \!=\!- 8 \pi G \uprho\nonumber\\
&&\qquad\qquad\qquad{63\over 2}\left[{1\over 2}\!
\left(\frac{d\varphi}{dr}\right)^2
\!\!e^{-2\theta} \!+\! V(\varphi) \right]=0.\eeq
Ref. \cite{Arsiwalla:2010bt} adopts the AdS${}_4$--Schwarzschild form of the radial metric component of the graviton star, 
\beq\label{adss}
\theta(r) = -\log\left[1 - {2GM(r) \over r}+ {r^2 \over \ell^2} \right],
\eeq
yielding the Misner--Sharp--Hernandez mass function, up to a factor of $4\pi$, 
\beq\label{mf}
\!\!\!\!\!\!\!\!\!\!\!\!{M(r)} \!=\!\! \int_0^r \!\!\!{\rm r}^{4}\! \left[\uprho({\rm r})
\!+\! {63\over 64 \pi G}\! \left(\frac{d\varphi({\rm r})}{d{\rm r}}\right)^{\!2}\!\! e^{-2\theta({\rm r})} \!+\! V(\varphi)\right]\!d{\rm r}.
\eeq
Besides, \clt{one can take the temporal and radial ({\footnotesize{$rr$}}) components of the equations of motion (\ref{jr1}) and subsequently add them, implying that} 
\beq\label{xir1}
\!\!\!\!\!\!\!\!\!\!\!\clt{\xi'(r) \!=\! \frac{63{\rm r}}{8}\!
\left(\!\frac{d\varphi({\rm r})}{d{\rm r}}\!\right)^{\!2} \!\!+\! 8\pi G (\uprho({\rm r})\!+\!\mathsf{p}({\rm r})) e^{2\theta({\rm r})}},
\eeq
\clt{which can be immediately integrated. Taking into account the equation of state for radiation (\ref{eosr}), the explicit form for the $\xi(r)$ function that composes the $4$-dimensional metric \eqref{ast1} reads }
\clt{\beq\label{xir}
\!\!\!\!\!\!\!\!\!\!\!\xi(r) \!&=&\!\frac12 \int_0^r\left[{63\over 4} \!
\left(\!\frac{d\varphi({\rm r})}{d{\rm r}}\!\right)^{\!2}\right.\nonumber\\&&\left.\;\;\qquad + \frac{44\pi G}5\!\left(1 \!-\! {2GM(r) \over r}\!+\! {r^2 \over \ell^2} \right)^{-2}\!\!\!\! \uprho({\rm r}) \!\right]{{\rm r}}\,d{\rm r}.
\eeq}
\!\!\clt{Therefore, handling the coupled system of equations (\ref{fi1}, \ref{mf}, \ref{xir}), with the AdS${}_4$--Schwarzschild form of the radial metric component of the graviton star (\ref{adss}) and the dilaton potential \eqref{pott}, yields the effective $4$-dimensional graviton star density, together with the conservation law $\nabla_\mu T^{\mu\nu}=0$, to read \cite{Arsiwalla:2010bt}
\begin{equation}
\label{eoseleven}
 \uprho(r) = e^{{81\over 2}\varphi(r) -
 11\xi(r)}\left(1 \!-\! {2GM(r) \over r}\!+\! {r^2 \over \ell^2} \right)^{-11}\,\uprho_{\scalebox{.56}{$0$}}\zeta,
\end{equation}}
\!where $\zeta=e^{-{63\over
 2}\varphi_{\scalebox{.56}{$0$}} +11\xi_{\scalebox{.56}{$0$}}}$ and $\uprho_{\scalebox{.56}{$0$}}=\displaystyle{\lim_{r\to0}}\uprho(r)$ is the density at the graviton star center. 
These equations can be used to find solutions numerically, using the shooting method, also with $\displaystyle{\lim_{r\to0}}\uprho'(r)=0$ \cite{Arsiwalla:2010bt}.
 The value $\xi_{\scalebox{.56}{$0$}}=\displaystyle{\lim_{r\to0}}\xi(r)$ at the star center can be derived by noticing that the equations of motion are invariant under a shift of $\xi(r)$. Also, by demanding an asymptotically AdS${}_4$
 metric, one must impose $\displaystyle{\lim_{r\to\infty}}\xi(r)=0$, which also constrains $\xi_{\scalebox{.56}{$0$}}$. Every central density $\uprho_{\scalebox{.56}{$0$}}$ determines the value $\varphi_{\scalebox{.56}{$0$}}=\displaystyle{\lim_{r\to0}}\,\varphi(r)$, when the limit $r\to0$ is taken in Eq. \eqref{eoseleven}. Besides, denoting by $M=\displaystyle{\lim_{r\to\infty}} {M}(r)$ in Eq. (\ref{mf}), the dilaton $\varphi(r)$ in the regime $r\gg 2M$ must correspond to a global minimum of the potential $V(\varphi)$ to yield an asymptotically AdS${}_4$ space \cite{Arsiwalla:2010bt}.

\section{DCE of AdS${}_4$ graviton stars}
 \label{2s}
 \clt{Configurational information measures have been employed to quantify the informational content and complexity of shape in field theories \cite{Gleiser:2018jpd}. The configurational entropy
(CE) and configurational complexity (CC), and their
continuum differential respective counterparts, DCE and DCC, play an important role as configurational information measures. 
The CE setup is influenced by Shannon's entropy \cite{Shannon:1948zz},
$
S = -\sum_{c\in \mathcal A}p_k\log p_k,$ 
stated for a set $\mathcal{A} = \{c_1, c_2,\ldots, c_N\}$ of symbols  with respective probability distribution $p_k = p(c_k)$.
The information content of a symbol, $I(c) = -\log_2 p(c)$, satisfies the expression 
\begin{equation}\label{ShannonInfo}
\langle I \rangle = \sum_{c\in\mathcal A} p(c)I(c) = -\sum_{c\in\mathcal A} p(c) \log_2 p(c) = S.
\end{equation}
Information can be then defined as the minimum number of bits necessary to encode a symbol to achieve a maximal transmission rate of messages along a channel. Therefore, the number of bits to lay up a specific symbol is $\log_2 N$.  
Unknownness in a uniform distribution can be described by $p(c) = 1/N$, whose entropy is then maximized at $\langle I \rangle = \log_2 N$. 
One can utilize base 2 and measure entropy
in bits, whereas physicists prefer base $e$ and measure entropy in nat\footnote{The natural unit of entropy.}. The CE has $\log |A|$ as an upper bound for equiprobable symbols and a null lower bound, when some symbol is certain.  
The CE encloses additional entropy formulations, namely the Hartley and the Nat entropies, respectively for $b=10$ and $b=e$, besides the Shannon's entropy for $b = 2$. Kolmogorov established the relationship between the CE and the Gibbs--Boltzmann thermodynamical entropy, 
${S_{\scalebox{.66}{\textsc{therm.}}}=-k_{{B} }\sum_i p_{i}\log p_{i}\,}$
where $k_B$ is the Boltzmann constant, and $p_i$ denotes the probability of a microstate. Also, the von Neumann entropy reads   
${S_{\scalebox{.66}{\textsc{Neumann}}}=-k_{{B} }\,{\rm {Tr}}(\mathring\rho \log \rho )\,}$, where $\mathring\rho$ is the density matrix.
Thermodynamics can be linked to information theory by the Boltzmann equation, 
$S_{\scalebox{.66}{\textsc{therm.}}}=k_{B}\log(W)$, 
where $W$ is the number of equiprobable microstates. Therefore, the CE measures how much information, necessary to determine a microstate, lacks. Thermodynamic entropy is proportional to the amount of Shannon information necessary to comprise isolated  microscopic states of a system \cite{Bernardini:2016hvx}. 
The Fourier transform of the energy density, 
\begin{equation}
\varrho(\omega)=\frac{1}{\sqrt{2\pi }}\int \;e^{i\omega r}\rho(r)\,dr,
\label{collectivecoordinates}
\end{equation}
is employed for emulating, in the continuum limit, the collective coordinates in statistical mechanics, namely, $\rho(r)\sim\sum e^{-i\omega_n r}\varrho(\omega_n)$. 
The normalized structure factor reads $
f(\omega_n)={\langle\;\left\vert \varrho(\omega_n)\right\vert ^{2}\;\rangle}/{\;s_n},$ where  $
s_n \sim \sum_{k=1}^n\langle\;\left\vert \varrho(\omega_k)\right\vert ^{2} \rangle\,,$ 
and quantifies the relative weight that each wave mode $\omega$ contributes. In the large  $n$ regime, the correlation is denoted by  
$
s=\lim_{n\to\infty}s_n$.  
It is worth to mention that in 
homogeneous hydrodynamical fluids, the collective coordinate average expected value vanishes, whereas the correlation of $f(\omega)=\lim_{n\to\infty}f(\omega_n)$ is not trivial. Despite of the averaged density being  constant in homogeneous hydrodynamical fluids, the energy density in general is allowed to fluctuate along configurations, capturing the system propensity to homogenization. }

 \clt{When the continuum limit is taken into account, the DCE can be  assumed ab initio, being here employed to scrutinize AdS${}_4$ graviton stars introduced in the previous section}. The DCE establishes a reliable procedure that quantifies the information content into the density. Denoting by $k$ the angular wavenumber, the Fourier transform 
\begin{eqnarray}
\uprho(k) =\frac{1}{\sqrt{2\pi }} \int_0^\infty \uprho(r)e^{-ikr}dr,\label{ftrans}
\end{eqnarray} defines the collective coordinates in the continuum limit, emulating the discrete case in statistical mechanics \cite{Bernardini:2016hvx}. Computational aspects are more viable when splitting Eq. \eqref{ftrans} into real and imaginary parts, \beq\uprho(k)=\uprho_{\scalebox{.66}{\textsc{re}}}(k)+i\uprho_{\scalebox{.66}{\textsc{im}}}(k),\eeq
 where 
\begin{subequations}
\begin{eqnarray}
\uprho_{\scalebox{.70}{\textsc{re}}}(k)&=& \int_0^\infty \uprho(r)\cos(kr)dr,\label{spl1}\\
\uprho_{\scalebox{.70}{\textsc{im}}}(k)&=& \int_0^\infty \uprho(r)\sin(kr)dr.\label{spl2}
\end{eqnarray}
\end{subequations}
Thus the modal fraction can be defined,  
\begin{eqnarray}\label{modall}
f_\uprho(k) = \frac{\uprho_{\scalebox{.70}{\textsc{re}}}^2(k)+\uprho_{\scalebox{.70}{\textsc{im}}}^2(k)}{\int_{0}^\infty \left[ \uprho_{\scalebox{.70}{\textsc{re}}}^2(\mathsf{k})+\uprho_{\scalebox{.70}{\textsc{im}}}^2(\mathsf{k})\right]\,d\mathsf{k}},\label{modalf}
\end{eqnarray}
and corresponds to the weight carried by the wave modes associated to the wavenumber $k$.  Therefore the modal fraction quantifies the contribution of wave modes to the power spectral density associated to the energy density. In fact, standing wave excitation modes can influence all the system components at a fixed frequency associated with these modes. Being the total energy finite, the power spectral density regarding the specific mode $k$, enclosed in the measure $dk$, is given by $
{P}\left(k\;\vert\; {d}k\right)\sim \left|\uprho(k)\right|^{2}{d}k$ \cite{Gleiser:2018kbq}, referring to the spectral energy distribution found in $dk$, also measuring how the density spatially fluctuates. The portion of information that is necessary to set out the shape of the density, here describing AdS${}_4$ graviton stars, is computed by the DCE, \clt{here equivalently taken as the differential configurational complexity \cite{Gleiser:2018jpd}}, defined by 
\cite{Gleiser:2012tu} 
\begin{eqnarray}
S[\uprho] = - \int_0^\infty\! \check{f}_{\uprho}(k)\log(\check{f}_{\uprho}(k))\,dk.\label{ce1}
\end{eqnarray}
where $\check{f}_\uprho(k)=f_\uprho(k)/f_\uprho^{\scalebox{.56}{\textsc{max}}}(k)$. 
In most cases the maximal value $f_\uprho^{\scalebox{.56}{\textsc{max}}}(k)$ is usually attained by the $k=0$ zero wave mode. It is worth mentioning that the uniform probability distribution makes the DCE as high as possible, since it evaluates a random probability distribution. 
The DCE is measured in natural units of information (nat), defined as the information content of an event, whose probability of occurrence is the inverse of the Euler number. Any spatially localized and Lebesgue-integrable function can be used for computing the DCE.

\begin{figure}[h]
\centering
\includegraphics[width=20.7pc]{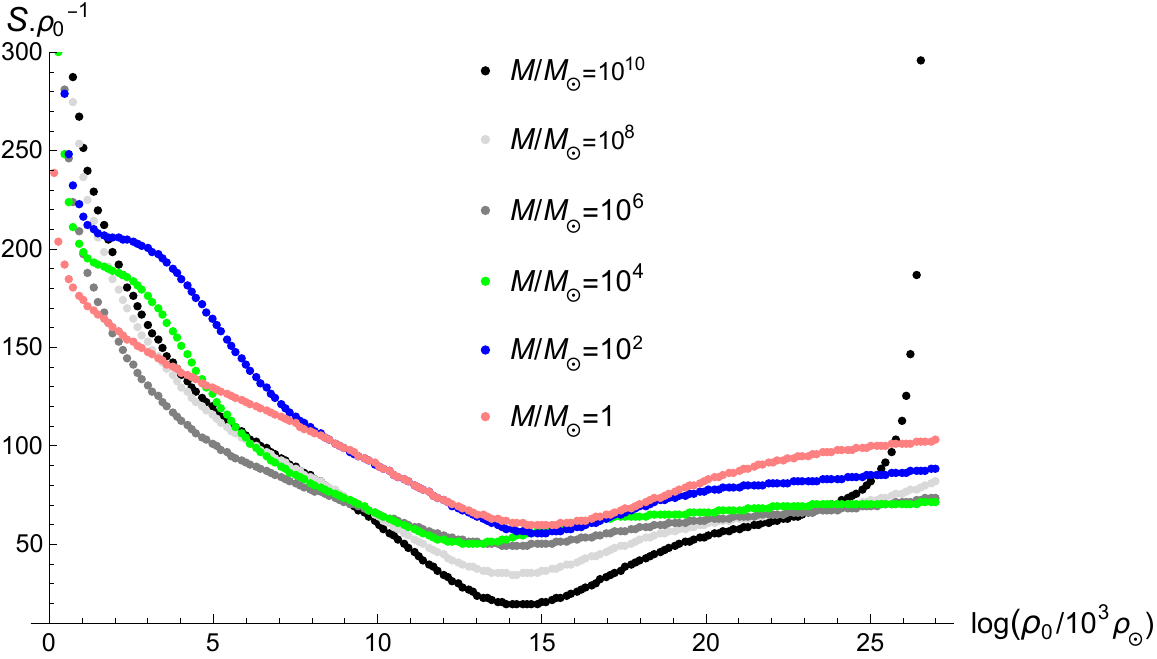}
\caption{\footnotesize{$S.\uprho_{\scalebox{.56}{$0$}}^{-1}$ of the AdS${}_4$ graviton star as a function of the logarithm of the star central density. The black plot shows the results for $M=10^{10}M_{\scalebox{.56}{$\odot$}}$, the light grey plot for $M=10^8M_{\scalebox{.56}{$\odot$}}$, the grey plot for $M=10^{6}M_{\scalebox{.56}{$\odot$}}$, the green plot for $M=10^{4}M_{\scalebox{.56}{$\odot$}}$, and the blue plot corresponds to $M=10^{2}M_{\scalebox{.56}{$\odot$}}$, whereas the case $M=M_{\scalebox{.56}{$\odot$}}$ is represented by the cinnabar plot.}}
\label{bs34}
\end{figure}

One can use therefore the AdS${}_4$ graviton star density \eqref{eoseleven} to calculate the DCE. After computing the Fourier transforms (\ref{spl1}, \ref{spl2}), the modal fraction \eqref{modalf} can be immediately derived. Subsequently, the DCE 
\eqref{ce1} can be then integrated, whose results are numerically derived and depicted in Fig. \ref{bs34}. In what follows, $M_{\scalebox{.56}{$\odot$}} = 1.9891 \times 10^{30}$ kg denotes the Solar mass and $\uprho_{\scalebox{.56}{$\odot$}}\approx 1.4 \times 10^4$ kg/m${}^3$ the Solar central density. Besides, in all the numerical analysis with data in Fig. \ref{bs34} and Table \ref{scalarmasses1}, the graviton star asymptotic mass value regards taking the limit $M=\displaystyle{\lim_{r\to\infty}} {M}(r)$ in Eq. (\ref{mf}). 
To establish a criterion of comparison to previous works \cite{Gleiser:2015rwa,Casadio:2016aum,Fernandes-Silva:2019fez}, the DCE is also here multiplied by the inverse of the central density to produce a quantity
that scales with dimensions of ${\rm mass}^{-1}$. Fig. \ref{bs34} shows that the DCE multiplied by the inverse of the central density, $S.\uprho_{\scalebox{.56}{$0$}}^{-1}$, has an absolute minimum value for the respective critical central density, for each value of mass. It suggests a critical configuration equilibrium value of the graviton star central density, also establishing a reliable range wherein AdS${}_4$ graviton stars are more stable from the configurational point of view. These minima are displayed in Table \ref{scalarmasses1}. The global critical values of the DCE in Fig. \ref{bs34} have, as absciss\ae\,, the critical central densities that are between 14 and 15 orders of magnitude the Solar density $\uprho_{\scalebox{.56}{$\odot$}}$, corresponding to neutron stars typical densities. This is shown in the third column of Table \ref{scalarmasses1}. 
The range of the central density chosen in Fig. \ref{bs34} is based on astrophysical observations, as very-low mass stars with masses below $\sim 0.5 M_{\scalebox{.56}{$\odot$}}$ are rarely observed and the most massive observed star has mass $1.65\times 10^5 M_{\scalebox{.56}{$\odot$}}$. 
Despite the maximum mass that a luminous accretor ultramassive black hole can reach is $\sim5 \times 10^{10}\,M_{\scalebox{.56}{$\odot$}}$, even larger ones have been theoretically predicted to have masses greater than $10^{12}\,M_{\scalebox{.56}{$\odot$}}$, however with no currently evidence still \cite{Carr:2020erq}. Therefore, the range of masses below covers all observed physical possibilities. It is also worth mentioning that data regarding observable stars regard mostly baryonic stars, whose nature is quite different, from the constitutive aspects, of the theoretically proposed graviton stars.
However, one must base on observations as starting points.
\begin{table}[h]
\begin{center}
\medbreak
\begin{tabular}{||c||c||c||}
\hline\hline \;$M$\; &\;$\left(S.\uprho_{\scalebox{.56}{$0$}}^{-1}, \log\left({\uprho_{\scalebox{.56}{$0$}}}/10^3{\uprho_{\scalebox{.56}{$\odot$}}}\right)_{\scalebox{.6}{min}}\right)$ \;&\;$\uprho_{\scalebox{.56}{$0$}}$\; \\\hline\hline
$M_{\scalebox{.56}{$\odot$}}$\;&\;$(15.418,\,61.648)$\;&\;$4.965\times 10^{15}$ $\uprho_{\scalebox{.56}{$\odot$}}$\\\hline
$10^2M_{\scalebox{.56}{$\odot$}}$\;&\;(15.379,\,58.553)\;&\;$4.775\times 10^{15}$ $\uprho_{\scalebox{.56}{$\odot$}}$\\\hline
$10^4M_{\scalebox{.56}{$\odot$}}$\;&\;(13.010,\,55.879)\;&\;$4.468\times 10^{14}$ $\uprho_{\scalebox{.56}{$\odot$}}$\\\hline
$10^6M_{\scalebox{.56}{$\odot$}}$\;&\;(14.018,\,49.548)\;&\;$1.224\times 10^{15}$ $\uprho_{\scalebox{.56}{$\odot$}}$ \\\hline
$10^8M_{\scalebox{.56}{$\odot$}}$\;&\;(14.215,\,33.875)\;&\;$1.491\times 10^{15}$ $\uprho_{\scalebox{.56}{$\odot$}}$\\\hline
$10^{10}M_{\scalebox{.56}{$\odot$}}$\;&\;(14.401,\,12.979)\;&\;$1.796\times 10^{15}$ $\uprho_{\scalebox{.56}{$\odot$}}$\\\hline\hline\hline\end{tabular}
\caption{Regarding data in Fig. \ref{bs34}. Second column: minima 2-tuples, where the absciss\ae\, regard the critical points of the graviton star central density $\log\left({\uprho_{\scalebox{.56}{$0$}}}/10^3{\uprho_{\scalebox{.56}{$\odot$}}}\right)_{\scalebox{.6}{min}}$ and the ordinates are the critical values of $S.\uprho_{\scalebox{.56}{$0$}}^{-1}$, for several values of the graviton star mass (first column). The third column shows the critical central densities corresponding to the respective global minima of the DCE.} \label{scalarmasses1}
\end{center}
\end{table}
\noindent
Polynomial functional interpolation forms of the DCE in Fig. \ref{bs34} is given in Appendix \ref{app}.

Critical points of the DCE do correspond to the most stable configurations, in the information entropy paradigm. States with higher information entropy, either need a higher amount of energy to be produced or are less frequently observed than higher DCE states. Those critical central densities of the AdS$_{4}$ graviton star in Table \ref{scalarmasses1} are center of insular domains that comprise stable configurations that best compress
 information into momentum space, consisting of the limit of almost lossless data compression. 
Analyzing Fig. \ref{bs34} reveals global critical values of the DCE, associated to critical points of the graviton star central density $\uprho_{\scalebox{.56}{$0$}}$. There is an absolute minimum for each value of $\uprho$, where the graviton star is more stable, from the point of view of the DCE. The DCE in Fig. \ref{bs34}, as a function of the graviton star mass $M$,
suggests that stars with lower masses are, in general, less stable from the point of view of the DCE. This again suggests a critical mass where the stellar system is configurationally more stable. 
Beyond these minima the DCE increases, in general, in a slower regime for the increment of the star mass, except for the specific case where $M=10^{10}M_{\scalebox{.56}{$\odot$}}$ in the range $\uprho_{\scalebox{.56}{$0$}}\gtrsim 1.49\times 10^{24}\uprho_{\scalebox{.56}{$\odot$}}$. 
The stability of 4-dimensional self-gravitating compact stars was previously scrutinized with the tools of the configurational entropy in Refs.~\cite{Gleiser:2015rwa,Casadio:2016aum,Fernandes-Silva:2019fez}, for Newtonian polytropes, neutron stars, and boson stars, including Bose--Einstein condensates of gravitons. The lower the DCE, the more the information, encoded into the wave modes that constitute the physical system, is condensed to portray the complexity of shape 
of the AdS$_{4}$ graviton star. 
 
\section{Concluding remarks}\label{conclu}

The DCE underlying AdS${}_4$ graviton stars was computed and analyzed, following the information protocol prescribed by the Eqs. (\ref{spl1}, \ref{spl2}, \ref{modalf}, \ref{ce1}).
The general profile of AdS${}_4$ graviton stars emulate the ones obtained for AdS${}_4$ radiation stars, seen as equilibrium configurations of self-gravitating massless radiation in asymptotically AdS space \cite{Arsiwalla:2010bt,Page:1985em}. As AdS${}_4$ graviton stars backreact to the surrounding geometric background, the ansatz (\ref{ansa}) makes the original Freund--Rubin spontaneous compactification AdS${}_4\times S^7$ metric to deform according to a dilaton field, also to take into account symmetric static configurations. Eqs. (\ref{jr1}, \ref{jr2}) were then derived as effective equations of motion in AdS${}_4$, whose solutions consist of the metric that describes AdS${}_4$ graviton stars, having the AdS${}_4$--Schwarzschild form for the radial component. The temporal component is then computed by Eq. (\ref{xir}) combined to Eq. (\ref{adss}), having mass function given by Eq. (\ref{mf}). 
 The star density (\ref{eoseleven}) is the main ingredient for computing the DCE, as it consists here of a spatially localized, Lebesgue-integrable scalar field. Analysis of the DCE, displayed in Fig. \ref{bs34} for several values of the star mass, indicates critical central densities of AdS${}_4$ graviton stars corresponding to the respective global minima values of the DCE, listed in Table \ref{scalarmasses1}. These critical central densities indicate physical states around which AdS${}_4$ graviton stars are configurationally more stable. The results here obtained comply with the existence of critical central densities in Newtonian polytropes, neutron stars, boson stars, and glueball stars, and stellar configurations formed of Bose--Einstein condensates of gravitons using the DCE 
 \cite{Gleiser:2011di,Gleiser:2015rwa,Casadio:2016aum,Fernandes-Silva:2019fez,daRocha:2017cxu}. Besides, following the results for AdS${}_4$ radiation stars \cite{Page:1985em}, Ref. \cite{Arsiwalla:2010bt} showed that there is a critical central density of AdS$_{4}$ graviton star that defines the Chandrasekhar stability range, whose limiting mass can be obtained formally from the Chandrasekhar equation by taking the limit of large central densities. Beyond these critical density ranges, AdS$_{4}$ graviton stars may collapse into AdS${}_4$--Schwarzschild black holes, that are in equilibrium with the associated Hawking radiation. Therefore the critical central density of AdS$_{4}$ graviton stars, corresponding to the respective global minima values of the DCE, corroborates this description. 
 
\clt{ One can still explore no-go theorems, emulating the results of Ref. \cite{Peng:2019uzw} to more general scenarios.}

\medbreak
\subsubsection*{Acknowledgments} I thank FAPESP (Grants No. 2017/18897-8 and No. 2021/01089-1) and the National Council for Scientific and Technological Development -- CNPq (Grants No. 303390/2019-0 and No. 406134/2018-9), for partial financial support. Also, to HECAP - ICTP, Trieste, for partial financial support and hospitality, and to Prof. K. Papadodimas for fruitful discussions.

\appendix
\section{Polynomial functional form of the DCE underlying AdS graviton stars}
\label{app}
The plots in Fig. \ref{bs34} were obtained from numerical computation of the Fourier transform of the density, the modal fraction, and subsequently the DCE, respectively using Eqs. (\ref{spl1}, \ref{spl2}, \ref{modall}, \ref{ce1}). Interpolation of all curves in Fig. \ref{bs34} yields a functional approximation within $0.4\%$ for the DCE\footnote{If one requires a smaller error, higher order interpolation polynomials can be employed, but here the use of quartic ones is enough for illustration.}, underlying AdS graviton stars, as a function of the central density $\uprho_{\scalebox{.56}{$0$}}$, for several values of mass. Explicit functional expressions for the DCE are respectively given by
\begin{widetext}
\begin{subequations}
\beq
\!\!\!\!\!\!\!\!\!\!\!\!\!\!\!\!{S}_{\scalebox{.66}{$M_{\scalebox{.66}{$\odot$}}$}}(\uprho_{\scalebox{.56}{$0$}})&=&\left[7.543\times 10^{-4}(\log\uprho_{\scalebox{.56}{$0$}})^4 -6.592 \times 10^{-2} (\log\uprho_{\scalebox{.56}{$0$}})^3+2.337
 (\log\uprho_{\scalebox{.56}{$0$}})^2-35.419 \log\uprho_{\scalebox{.56}{$0$}} +259.138\right] \uprho_{\scalebox{.56}{$0$}},\label{s1}\\
 \!\!\!\!\!\!\!\!\!\!\!\!\!\!\!\!{S}_{\scalebox{.66}{$10^2M_{\scalebox{.66}{$\odot$}}$}}(\uprho_{\scalebox{.56}{$0$}})&=&\left[1.191\times 10^{-3}(\log\uprho_{\scalebox{.56}{$0$}})^4 -0.744 (\log\uprho_{\scalebox{.56}{$0$}})^3+16.361
 (\log\uprho_{\scalebox{.56}{$0$}})^2-148.887 \log\uprho_{\scalebox{.56}{$0$}} +549.248\right] \uprho_{\scalebox{.56}{$0$}},\label{s2}\\
 \!\!\!\!\!\!\!\!\!\!\!\!\!\!\!\!{S}_{\scalebox{.66}{$10^4M_{\scalebox{.66}{$\odot$}}$}}(\uprho_{\scalebox{.56}{$0$}})&=&\left[8.140\times 10^{-3} (\log\uprho_{\scalebox{.56}{$0$}})^4 -0.533 (\log\uprho_{\scalebox{.56}{$0$}})^3+12.401
 (\log\uprho_{\scalebox{.56}{$0$}})^2-119.608 \log\uprho_{\scalebox{.56}{$0$}} +462.309\right] \uprho_{\scalebox{.56}{$0$}},\label{s3}\\
 \!\!\!\!\!\!\!\!\!\!\!\!\!\!\!\!{S}_{\scalebox{.66}{$10^6M_{\scalebox{.66}{$\odot$}}$}}(\uprho_{\scalebox{.56}{$0$}})&=&\left[1.239\times 10^{-3}(\log\uprho_{\scalebox{.56}{$0$}})^4 -0.775 (\log\uprho_{\scalebox{.56}{$0$}})^3+16.922
 (\log\uprho_{\scalebox{.56}{$0$}})^2-149.329\log\uprho_{\scalebox{.56}{$0$}} +498.935\right] \uprho_{\scalebox{.56}{$0$}},\label{s4}\\
 \!\!\!\!\!\!\!\!\!\!\!\!\!\!\!\!{S}_{\scalebox{.66}{$10^8M_{\scalebox{.66}{$\odot$}}$}}(\uprho_{\scalebox{.56}{$0$}})&=&\left[1.016\times 10^{-2}(\log\uprho_{\scalebox{.56}{$0$}})^4 -0.633 (\log\uprho_{\scalebox{.56}{$0$}})^3+13.804
 (\log\uprho_{\scalebox{.56}{$0$}})^2-122.073 \log\uprho_{\scalebox{.56}{$0$}} +422.560\right] \uprho_{\scalebox{.56}{$0$}},\label{s5}\\
 \!\!\!\!\!\!\!\!\!\!\!\!\!\!\!\!{S}_{\scalebox{.66}{$10^{10}M_{\scalebox{.66}{$\odot$}}$}}(\uprho_{\scalebox{.56}{$0$}})&=&\left[2.395\times 10^{-2}(\log\uprho_{\scalebox{.56}{$0$}})^4 -1.468 (\log\uprho_{\scalebox{.56}{$0$}})^3+31.075
 (\log\uprho_{\scalebox{.56}{$0$}})^2-261.644 \log\uprho_{\scalebox{.56}{$0$}} +767.404\right] \uprho_{\scalebox{.56}{$0$}},\label{s6}
\eeq
\end{subequations}
\end{widetext}

\end{document}